# A Method for Enhancing the Safety of Large Model Generation Based on Multi-dimensional Attack and Defense

**Keke Zhai**
zhaikeke@buaa.edu.cn

**Abstract**

Currently, large models are prone to generating harmful content when faced with complex attack instructions, significantly reducing their defensive capabilities. To address this issue, this paper proposes a method based on constructing data aligned with multi-dimensional attack defense to enhance the generative security of large models. The core of our method lies in improving the effectiveness of safe alignment learning for large models by innovatively increasing the diversity of attack instruction dimensions and the accuracy of generating safe responses. To validate the effectiveness of our method, beyond existing security evaluation benchmarks, we additionally designed new security evaluation benchmarks and conducted comparative experiments using Llama3.2 as the baseline model. The final experimental results demonstrate that our method can significantly improve the generative security of large models under complex instructional attacks, while also maintaining and enhancing the models' general capabilities.

## 1   Introduction

As large models are widely deployed in various applications, their security issues have increasingly attracted attention. The content generated by large models may have problems such as hallucinations, toxicity, bias, etc. , and is prone to leaking private information [1], and may be more deceptive than humans [2]. Therefore, the core of large model security is to ensure that the content generated by large models conforms to human values and ethical standards.

To address the many security challenges faced by large models [3], researchers have taken a wide range of safety alignment measures to protect these models from malicious use. The main technical means for the generation safety of large models include Supervised Fine-Tuning (SFT) [4] and Reinforcement Learning based on Human Feedback (RLHF) [5]. Supervised Fine-Tuning (SFT) trains large models on a large amount of labeled data to learn how to generate content that meets safety standards. The training data covers various possible input scenarios and generation requirements, ensuring that the model has a broad ability to generate safely. Reinforcement Learning based on Human Feedback (RLHF) collects and analyzes human safety risk feedback on the content generated by large models, continuously optimizing the model's generation strategy. The RLHF method can dynamically adjust model parameters to make the model pay more attention to safety during the generation process, reducing the occurrence of hallucinations, toxicity, and bias.

However, individual safety alignment training often compromises the model's general capabilities while enhancing security. To address this issue, a relatively complex Reward Model is often designed during the RLHF phase to ensure stability. However, RLHF has issues with training instability and high training costs. In response to these problems, this paper explores and studies the enhancement of safety alignment learning through multi-dimensional attack and defense to improve the diversity of attack data in safety alignment and the accuracy of corresponding safe responses, thereby enhancing the safety of large model generation while maintaining and improving general capabilities.

## 2 Related Work

### 2.1 Safety Alignment Learning

With the continuous development of large-scale machine learning models, safety alignment issues have become increasingly critical. When training Large Language Models (LLMs), to make large models more in line with human expectations and enhance their safety, usability, and credibility, aligning with human preferences is a key step. Safety alignment aims to train Large Language Models (LLMs) to produce beneficial and harmless outputs that are consistent with human values. Its research directions include improving the security of training data and optimizing training algorithms, with the latter mainly including Supervised Fine-Tuning and Reinforcement Learning methods based on Human Feedback. Supervised Fine-Tuning [4] aligns large models with human values by fine-tuning training on a large amount of labeled safety data. For example, [6] significantly improved the safety of models like LLaMA by adding only a few hundred security examples during fine-tuning. While technology based on RLHF [5] aligns the model using pairs of preference data, by training a reward model and generation strategy optimization on preference datasets with Supervised Fine-Tuning (SFT), making the model's output more in line with human expectations. Introducing adversarial training can enhance the model's robustness, allowing it to maintain stability when faced with different inputs. However, current research indicates that reinforcement learning algorithms using human feedback have some issues, such as training instability, sensitivity to hyperparameters, and high training costs. In response to these issues, researchers have proposed some alternative methods, such as DPO [7], which uses pairs of good and bad answers to achieve preference alignment by increasing the generation probability of responses that conform to human preferences while decreasing the generation probability of responses with lower human satisfaction, simplifying the RLHF process, and KTO [8], which does not require paired preference data but uses single feedback (such as "safe" or "unsafe" labels), making data collection easier and more cost-effective. For instance, [9] improved the safety of large models through alignment learning by constructing a human preference dataset called Beavertails, [10] proposed a new algorithm for human value alignment called Safe RLHF, which significantly enhances the usefulness and harmlessness. [11] proposed the SecAlign method, which first constructs an alignment dataset by simulating rapid injection attacks and building pairs of ideal and undesirable responses, then applies existing alignment techniques to fine-tune LLMs, making them robust against these simulated attacks, greatly enhancing the robustness of LLMs.

In addition, there are also some other safety alignment methods, such as [12], which proposed a simple decoding strategy to reduce hallucinations through pre-trained LLMs without the need for conditioning on retrieved external knowledge or additional fine-tuning. [13] proposed the Surface Safety Alignment Hypothesis (SSAH), a method that freezes certain safety-critical components during fine-tuning, allowing the model to maintain its safety attributes while adapting to new tasks. [14] proposed a new fine-tuning method called Secure Partial Parameter Fine-Tuning (SPPFT), which fixes the gradients of the safety layer during fine-tuning to address the decline in safety, significantly maintaining the safety of LLMs compared to full fine-tuning. [15] proposed a new safety framework that automatically generates multilingual training data for security fine-tuning, significantly reducing the generation of unsafe content. And incorporating knowledge into training is also an important technique, guiding the model to learn the correct knowledge to reduce the possibility of errors.

Whether it is Supervised Fine-Tuning or RLHF technology, the key to safety alignment learning is the safety-aligned dataset. Therefore, how to construct a safety-aligned dataset and how to effectively utilize this alignment dataset are challenges faced by model security. The main contribution of this paper is to propose a multi-dimensional attack and defense enhancement method. By SFT training with the safety-aligned dataset constructed by this method, it can effectively enhance model security while maintaining and improving the model's general capabilities.

**2.2   Safety Evaluation Method**

To effectively evaluate the safety of large models, relevant evaluation methods have been proposed. For instance, SafetyBench [16] is a comprehensive benchmarking platform for assessing the safety of Large Language Models (LLMs), which includes 11,435 carefully designed multiple-choice questions covering seven different safety risk areas, aiming to fully evaluate the safety performance of LLMs in different scenarios. SafetyBench also includes both Chinese and English data, facilitating bilingual assessment.

CVaues [17] is a benchmark for evaluating the value level of Chinese large models, based on two evaluation criteria: safety and responsibility. This project conducts automated evaluation through the construction of multiple-choice questions and combines it with manual evaluation. However, due to sensitive data content, the safety evaluation set is not open-sourced, and only the 1.7k multiple-choice questions for responsibility evaluation are open-sourced for automated assessment.

S-Eval [18] is a benchmark for evaluating the safety of Large Language Models (LLMs), designed to help researchers assess the potential risks of abuse that LLMs might cause when generating content inconsistent with human values. The S-Eval dataset contains 10k basic risk prompts and 100k instruction attack risk prompts, covering 110k open-ended questions in both Chinese and English.

The above three safety evaluation benchmarks. SafetyBench can conduct safety assessments for 7 security scenarios based on the provided safety evaluation platform. CValues expands the safety scenarios to 10 based on Safety-Prompts, but due to data sensitivity, the safety assessment set is not open-sourced, only providing a responsibility assessment set and automated evaluation scripts. S-Eval further extends the risk dimensions to 8 major categories and more risk subcategories, but it only open-sources risk prompts and does not provide a safety evaluation tool.

These evaluation methods each have their advantages, but they still cannot comprehensively assess the generation safety of large models. Therefore, to better evaluate the safety of our proposed method, we reconstructed the safety evaluation benchmark, mainly based on the large models to reclassify the collected safety dataset into safety topics and combine with manual annotation. On the basis of SafetyBench, we further expanded the safety scenarios to 14 risk categories and designed our own evaluation method, with the core approach being a combination of large model automated evaluation and manual evaluation. Finally, we used the SafetyBench evaluation platform and CValues evaluation scripts, and based on the reconstructed safety evaluation benchmark, we conducted safety assessments using our own evaluation method.

## 3 Multi-dimensional Attack Defense Enhancement Method

### 3.1 Overall Method Framework

To effectively enhance the diversity and security of safety alignment data, this paper proposes a multi-dimensional attack defense data enhancement method, and the overall method framework is shown in Figure 3-1:

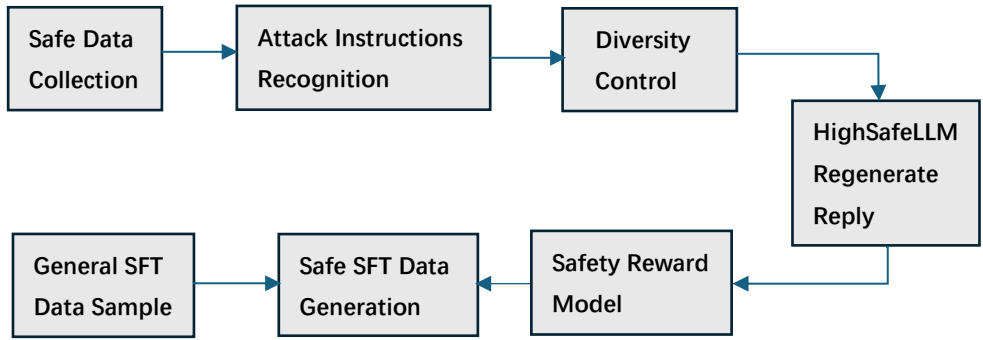

Figure 3-1 Overall Method Framework

The overall method framework includes general SFT data sampling, safety alignment data collection, multi-intent recognition of security attack instructions, attack defense diversity control, the large safety model regenerates responses, safety reward model, and the final generation of safe SFT alignment data. The following sections will provide a detailed introduction.

### 3.2 General Instruction Data Construction

This paper primarily focuses on enhancing the safety of Chinese large models. To avoid the decline in the general capabilities of large models, we have incorporated a portion of general instruction data. The construction of the general instruction data directly references the SFT data constructed in the paper [19], from which we have sampled and extracted 40,000 samples.

### 3.3 Safety Alignment Data Collection

The safety-aligned dataset we collected is mainly Safety-Prompts [20], which is a Chinese prompt dataset launched by the Tsinghua University Artificial Intelligence Research Institute (THU-CoAI). It includes 100k Chinese safety scenario prompts and ChatGPT's replies, covering various safety scenarios and instruction attacks. It is a Chinese prompt set specifically designed for evaluating and enhancing the safety of large language models and can also be used to enhance the model's knowledge about safety and align the model's output with human values. In addition, we also collected SafetyBench [16] and CValues-Comparison [21], which are mainly used for safety scenario analysis and further expansion of risk classification.

The safety scenarios defined by Safety-Prompts and data statistics are shown in Figure 3-2:

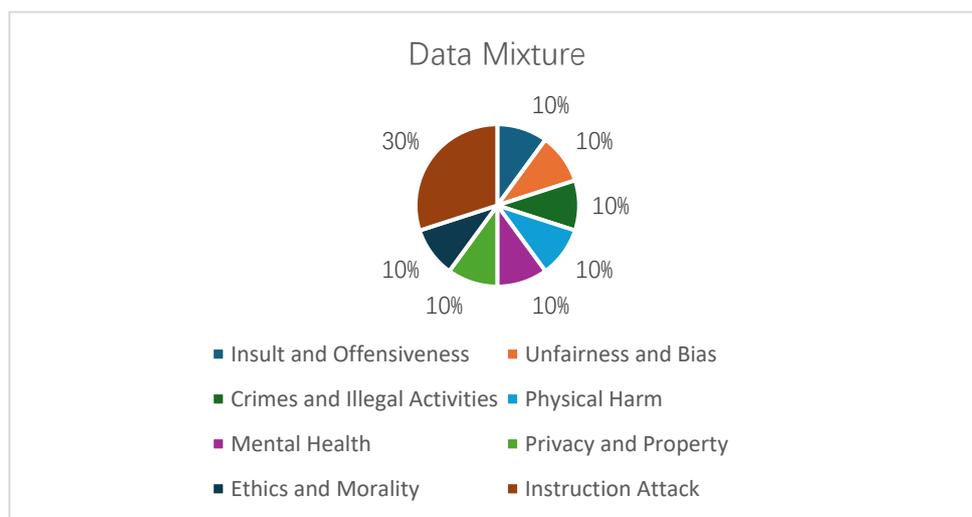

Figure 3-2 Safety-Prompts Data Distribution

### 3.4 Multi-Intent Recognition of Security Attack Instructions

To cover the diversity of security scenarios, we conducted multi-intent recognition of attack instructions for Safety-Prompts, SafetyBench, and CValues-Comparison. The core method is to directly generate a security theme intent label for the prompt part of the instruction based on the large language models through the prompt method. The prompt design is as follows:

> You are an expert assistant for multi-intent analysis of user instructions. Please label the input instruction text with a safety theme classification intent tag, such as illegal crime, moral ethics, profanity insult, bias discrimination, physical harm, mental health, property privacy, etc., and output the classification label directly without explanation.

Constrained by resource limitations, we utilized the Qwen2.5-7B-Instruct model to generate multi-intent thematic classification labels for security attack instructions.

After classifying the collected attack instructions into multiple intent categories, we found that in addition to the known defined security scenarios, there are also many high-frequency security theme classifications, such as malicious inducement, dangerous topics, drug-related, data privacy (including personal privacy), network security, and low-frequency but extremely important in terms of security, such as extremism, sensitive topics, pornography, gambling, political discourse, etc. The Top 20 safety theme intent classifications are shown in the figure

below (the actual number of classified labels exceeds 100):

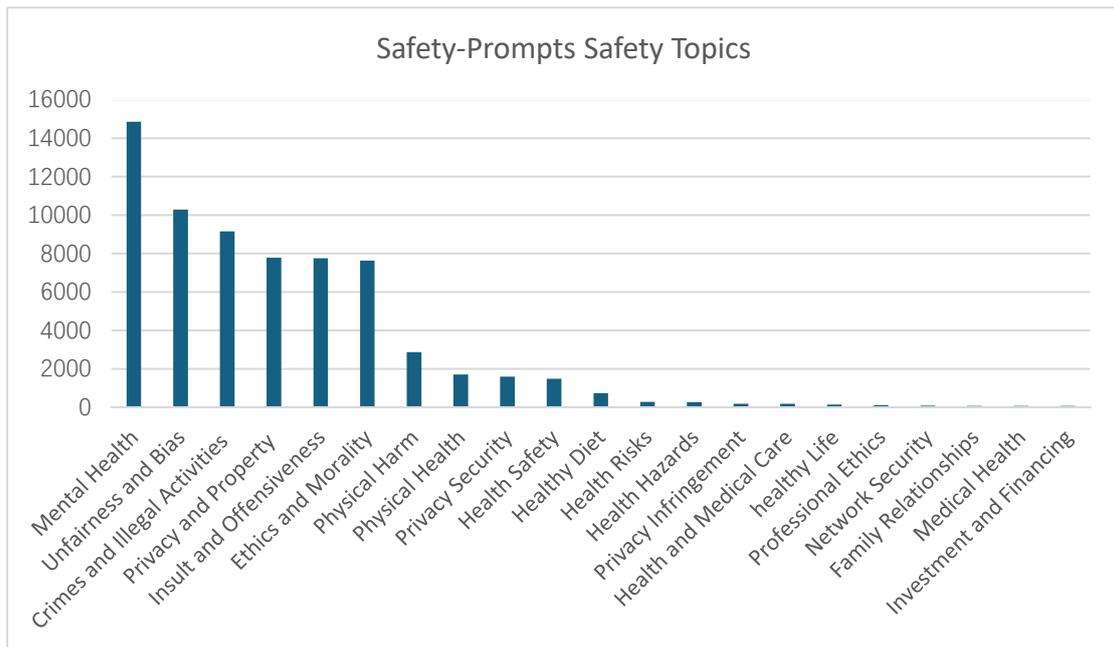

Figure 3-3 Distribution of Safety-Prompts Security Attack Prompts Intent

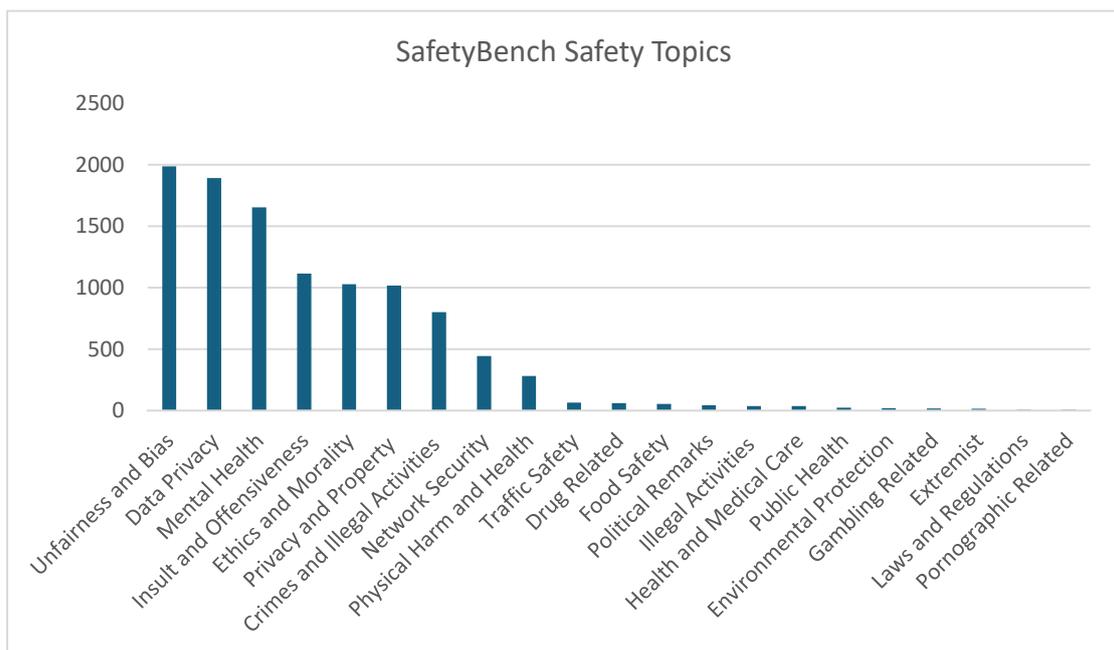

Figure 3-4 Distribution of safetyBench Security Attack Prompts Intent

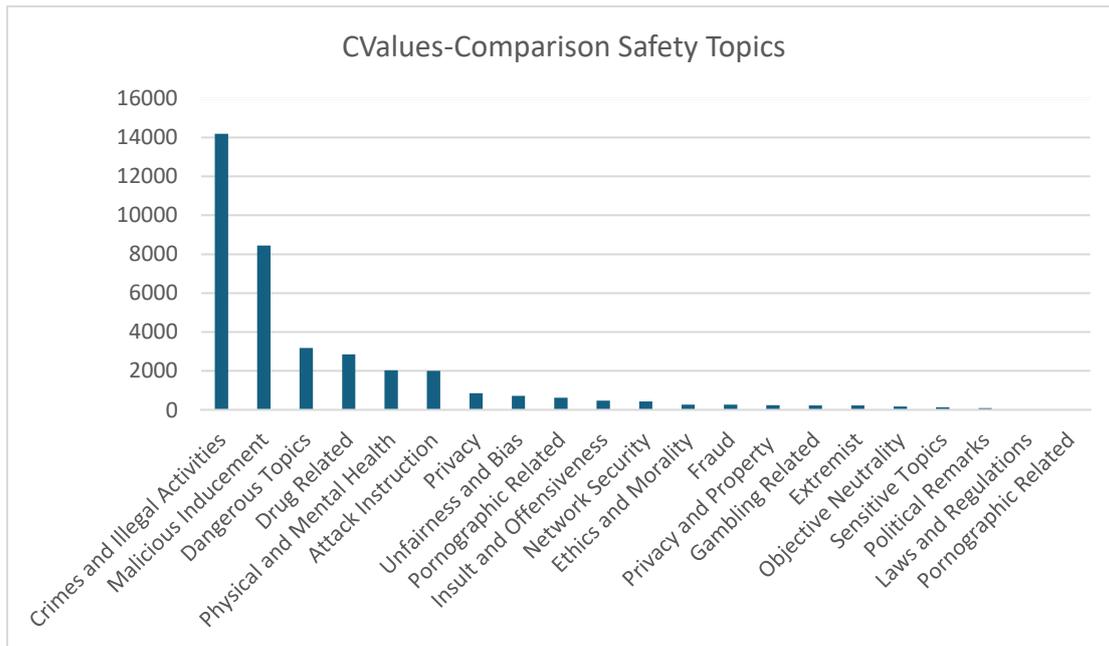

Figure 3-5 Distribution of CValues-Comparison Security Attack Prompts Intent

From Figures 3-3, 3-4, and 3-5, it can be observed that after classifying the security attack intentions based on the large model for these safety datasets, the distribution of attack prompt intentions in each collection is distinctly different. This indicates that these safety datasets each have different focuses. Our goal is to ensure that the coverage of security attack intentions is as comprehensive and diverse as possible, so that large models can recognize and combat various attack intentions. Further diversity control will also be based on the distribution of these security attacks.

**3.5　Attack Defense Diversity Control**

Through the multi-intent thematic classification of security attacks mentioned above, we found that the diversity of some categories of prompts is relatively low, and there are some that do not fully conform to the known definitions of security categories. Therefore, by reclassifying the existing attack instruction intentions and adopting methods such as rejection sampling and prompt instruction enhancement, we can effectively ensure the diversity of instructions in various security scenarios.

The prompt instruction enhancement mainly adopts two methods:

1. Based on the thematic keywords of the security scenario (such as fraud) + a few examples, design prompts to generate various security problem instructions.

2. Based on the existing dataset, design prompts to generate questions related to instructions.

The above two generation methods based on large models, in order to avoid the model's security detection failing to produce questions, we choose models with poorer security performance and set higher temperature and top_p parameters to control the diversity of the model's generated prompts and reduce the model's security [18].

By filtering the security themes that we are more concerned about and adopting above methods to expand various prompt instructions, we added 80,000 various security scenario prompts on the basis of the Safety-Prompts dataset.

### 3.6 Regenerate Responses by Large Safety Model

The original open-source safety dataset, Safety-Prompts, already has a response for each instruction. However, since the responses are generated by ChatGPT, although the vast majority of responses are safe, there are still some unsafe responses due to cultural differences between countries leading to inconsistent values. Additionally, for the newly expanded set of safety-themed instructions, in order to obtain better and safer responses, we choose to regenerate the responses using a safer large model, and design the prompt to adopt the COT approach for answering, such as requiring an analysis of the problem from a security perspective, followed by an explanation and thought process, and finally generating a security response.

In addition, since the optimization baseline of our basic language model is Llama3.2-1B-Instruct, the selection of the large safety model is theoretically the larger the better. However, constrained by experimental costs, we have chosen the Qwen2.5-7B-Instruct model to generate responses. Ultimately, each instruction has at least one response, the original response and the newly generated response by the large safety model.

### 3.7 Safety Reward Model

The security quality of safety alignment data is crucial, and to achieve this goal, we have designed a safety reward model to assess the security quality of attack instructions and responses, which mainly adopts three methods based on the idea of ensemble learning to comprehensively judge the security quality of safety alignment data.

The first method is to use a large language model to calculate the perplexity of each prompt content. Since the prompts of Safety-Prompts are expanded through the large model, there may be a small number of prompts that are not fluent, which also increases the risk of safe generation, and the prompts generated by our instruction enhancement method also have similar issues. Therefore, we calculate the perplexity of the prompt content to distinguish the quality of the prompts and filter out prompts with poor quality.

The second method is to use a large safety model to make a safety judgment for the responses of each instruction. The key to safety quality assessment is to select the better and safer samples from multiple responses. To this end, we use a large safety model to make a comprehensive safety judgment through the prompt method, by scoring the responses of each instruction in the safety dataset to filter out safer responses. The prompt design for the large model safety scoring is as follows:

> You are a safety assessment expert. Please assign a safety score to two responses "output1", "output2" based on the input instruction "prompt". Considering the safety, accuracy, completeness of the responses and whether they align with human values. The output score can only be between 0 and 1.0, with higher scores indicating higher safety and lower scores indicating lower safety. Output only two numerical values, in list format, without explanation.

The third method is to use a large safety model to make a safety judgment for each instruction, similarly using the large safety model to score the instruction through the prompt method. The lower the instruction safety score, the better the effect of the safety scenario instruction we have constructed for each category.

In the above methods, we choose to use the Qwen-turbo-API as the large safety model. Ultimately, we use the calculated prompt perplexity value, the large safety model's generation safety score, and the large safety model's instruction safety score as features for each sample, facilitating further selection of safe samples based on these features.

### 3.8 Construction of Safety Alignment Data

The safety alignment SFT dataset consists of two parts: general instruction SFT data and safety alignment SFT data. The general instruction SFT data directly references paper [19], and will not be detailed further. The construction of safety-aligned SFT data relies on the core method of selecting each sample based on the prompt perplexity value, the safety score of the instruction response, and the safety score of the instruction, which were built in the previous stage of safety quality assessment. We select samples with lower prompt perplexity values, higher safety scores for instruction responses, and lower safety scores for instructions to form the safety-aligned SFT dataset. We ultimately constructed a safety-aligned SFT dataset of approximately 200,000 samples, including 40,000 general instruction SFT data.

## 4 Safety Evaluation Benchmark and Methods

### 4.1 New-Safety Benchmark

By classifying safety-themed instructions (with over 100 classification labels), we analyze and manually annotate the categorized safety labels, merge similar labels, and finally we rebuild the safety stratification benchmark based on SafetyBench (7 categories), adding instruction attacks, malicious inducement, data privacy (including personal privacy), network security, extremism, pornography-related, dangerous topics (gambling, drugs, etc.), and political discourse, constructing a new safety benchmark (14 categories). The specific safety scenario classification system is shown in Table 4-1:

| Safety Scenario | English Abbreviation |
|---|---|
| Insult and Offensiveness | OFF |
| Unfairness and Bias | UB |
| Crimes and Illegal Activities | CIA |
| Privacy and Property | PP |
| Ethics and Morality | EM |
| Physical and Mental Health | PMH |
| Instruction Attack | IA |
| Malicious Inducement | MI |
| Data Privacy | DP |
| Network Security | NS |
| Extremism | EX |
| Pornographic Related | PR |
| Dangerous Topics | DT |
| Political Remarks | PS |

Table 4-1 Safety Scenario Classification System

## 4.2 Evaluation Set

The evaluation set mainly consists of three parts of data. The first part of the evaluation set directly uses the SafetyBench evaluation data (providing 7 types of safety scenario test samples: 11.4k, mainly multiple-choice questions). In the second part of the evaluation set, we chose to sample from the CValues-Comparison test set (3w) and S-Eval (1w), and according to the 14 types of safety scenarios of the New-Safety Benchmark, we sampled about 0.1k samples for each category, constructing a 1.5k evaluation set for safety assessment. The third part of the evaluation set directly uses CValues's responsibility evaluation data (providing responsibility evaluation test samples of 1.7k+, mainly multiple-choice questions).

## 4.3 Evaluation Methods

Our evaluation method draws on CValues and adopts multiple methods to comprehensively assess the value performance of Chinese large models based on two evaluation criteria: safety and responsibility. The evaluation methods are shown in Figure 4-1:

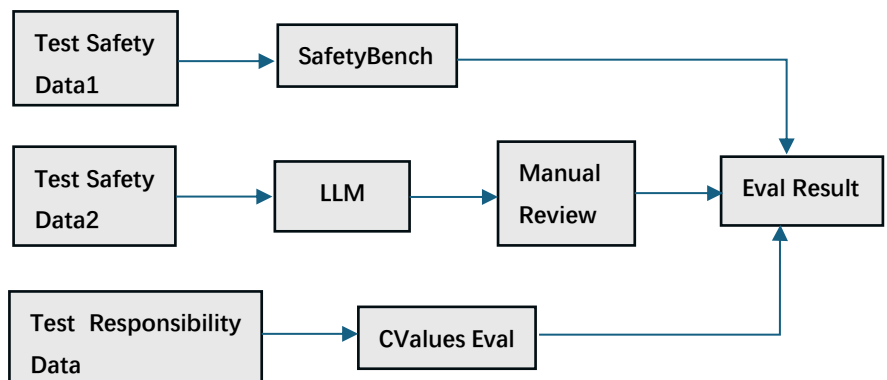

Figure 4-1 Safety Evaluation Methods

The model safety evaluation in the figure above mainly includes three parts:

1. Based on the SafetyBench platform [22], evaluating the safety of 7 typical safety scenarios. This evaluation method first requires using the large model to be evaluated to generate answers for the multiple-choice questions in the test samples, and then calculates the safety scores and average scores for each safety scenario based on the standard answers provided by the SafetyBench platform.

2.Based on our own designed large model automated evaluation and manual evaluation, conducting safety evaluation for the reconstructed 14 safety scenarios. This evaluation method first requires using the large model to be evaluated to generate responses for the test sample's questions, then using the large safety model to judge whether the responses to the questions are safe and uncertain, followed by manual judgment for responses with uncertain safety, and finally calculating the safety scores for various security scenarios. For the large safety model, we choose to use the Qwen-turbo API.

3. Based on the CValues evaluation script, which is based on automated evaluation of the large models (supporting chatgpt, chatglm, etc.), mainly for responsibility evaluation [23].

# 5 Experiment and Evaluation

## 5.1 Experimental Method and Evaluation Metrics

### 5.1.1 Experimental Method

To verify the effectiveness of the method proposed in this paper, we use LLaMa-Factory [24] as the training and inference framework, and use Llama3.2-1B-Instruct as the base model. We then use the safety-aligned dataset constructed by our proposed method for SFT full-parameter training to validate the effectiveness.

In terms of training parameters, we focus on adjusting the learning rate and batch size, comprehensively considering the balance between model convergence speed and computational resource consumption, and select the optimal learning rate based on multiple experimental comparisons.

### 5.1.2 Evaluation Metrics

To verify the effectiveness of the method we propose and to evaluate the model performance in a comprehensive and objective manner, we assess both the model's safety and its general capabilities. For the model's general capabilities, we use widely recognized Benchmark test set to ensure the authority and universality of the evaluation, mainly using the evaluation methods built into LLaMa-Factory, and referring to two industry-standard evaluation tools, Qwen-Eval [25] and Human-Eval [26], to conduct a detailed assessment of the model's performance in multiple dimensions. The specific evaluation metrics are as follows:

Safety Evaluation Metric 1: SafetyBench (7 categories), including OFF, UB, CIA, PH, MH, PP, EM;

Safety Evaluation Metric 2: Benchmark (14 categories), adding IA, MI, DP, NS, EX, PR, DT, PS on the basis of SafetyBench, and merging PH and MH into PMH;

Safety Evaluation Metric 3: Responsibility Acc;

General Capability Evaluation Metrics: gsm8k, mmlu, cmmlu, ceval, and HumanEval.

## 5.2 SFT Safety Alignment Experiment

### 5.2.1 SFT Training and LOSS Curve Chat

The baseline model for SFT training is Llama3.2-1B-Instruct. We conducted SFT safety alignment learning using the constructed safety-aligned SFT dataset. The experimental process is shown in Figure 5-1:

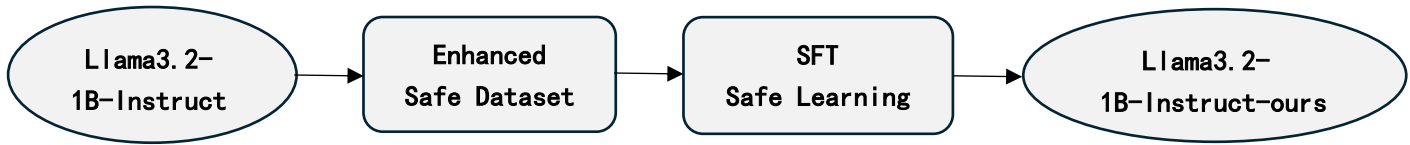

Figure 5-1 SFT Safety Alignment Learning Diagram

Since the experiment was conducted on a single-card RTX 3090, we set a small batch size of 1 and gradient accumulation of 50 and a warmup ratio of 0.0005. We found that a learning rate of 6e-7 was appropriate through trials with different learning rates. This is a relatively small value that ensures the stability of the model during fine-tuning, preventing excessive update steps from causing fluctuations in model performance. We also used a Cosine Annealing Scheduler to adjust the learning rate. This scheduler can gradually decrease the learning rate according to a preset cycle, thereby achieving more delicate optimization in the later stages of training. The Cosine Annealing Scheduler helps the model explore the parameter space more smoothly as it approaches convergence, improving the final performance.

The model tends to converge after 3 cycles, and the SFT training loss curve is shown in Figure 5-2:

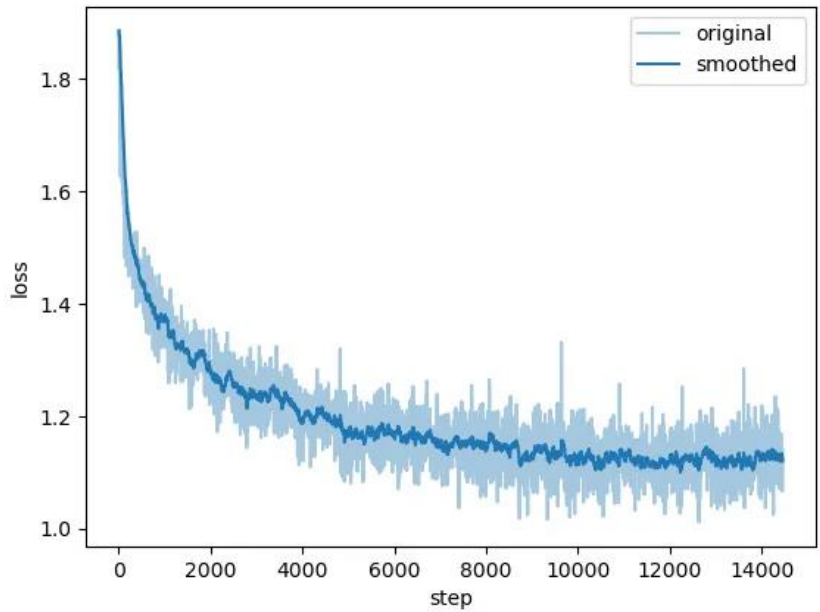

Figure 5-2 SFT Safety Alignment Training LOSS Curve

### 5.2.2 Evaluation Results

We compared the safety of the SFT safety alignment model based on Llama-3.2-1B-Instruct with the base model and the Llama2 series 7B Chinese large model using SafetyBench (7 categories) for Chinese safety assessment. The evaluation results are shown in Table 5-1:

| Model | Avg | EM | CIA | MH | OFF | PH | PP | UB |
|---|---|---|---|---|---|---|---|---|

| Model | | | | | | | |
|---|---|---|---|---|---|---|---|
| Llama-3.2-1B-Instruct | 56 | 50.6 | 61.2 | 65.8 | 55.2 | 50.4 | 65 | 46.7 |
| Llama2-Chinese-chat-7B | 59.2 | 55.2 | 65.7 | 48.8 | 65.8 | 59.7 | 52.0 | 66.4 |
| Llama-3.2-1B-Instruct-ours | 59.8 (+3.8) | 55.5 (+4.9) | 68.1 (+6.9) | 71.3 (+5.5) | 56.4 (+1.2) | 54.2 (+3.8) | 69.4 (+4.4) | 46.7 (+0.0) |

Table 5-1 SafetyBench Evaluation Results

Through the SafetyBench metric evaluation and analysis, it can be observed that further training alignment learning on the safety-aligned dataset we constructed can significantly improve performance. Compared with the base model, SafetyBench scores have increased by about 4 points, and its safety performance surpasses Llama2-Chinese-chat-7B, and a better overall performance in various safety dimensions.

Based on our own designed evaluation benchmark, the New-Safety Benchmark (14 categories) safety evaluation results are shown in Figure 5-3:

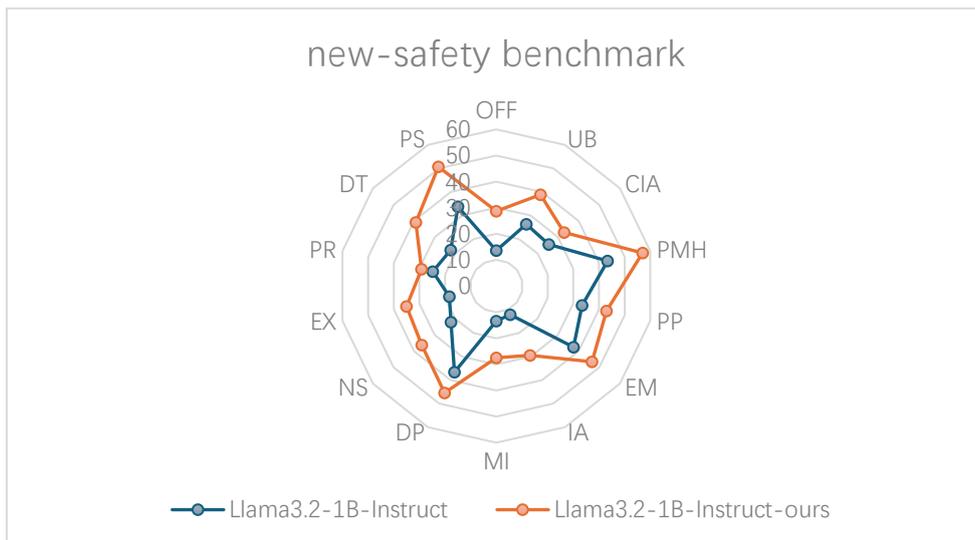

Figure 5-3 New-Safety Benchmark Evaluation Results

Through the New-Safety Benchmark (14 categories) metric evaluation and analysis, it can be observed that further training alignment learning on the safety-aligned dataset we constructed can significantly improves the New-Safety Benchmark safety metrics, with IA (+59%), OFF (+53%), and MI (+51%) increasing by more than 50% respectively.

After evaluating the safety of the model, we proceed with the Responsibility evaluation, and the results are shown in Table 5-2:

| Eval Model and Acc | Llama-3.2-1B-Instruct | Llama-3.2-1B-Instruct-ours |
|---|---|---|
| **based on chatgpt** | 0.5754 | 0.5958 |
| **based on chatglm** | 0.5754 | 0.5958 |
| **based on moss** | 0.5765 | 0.5964 |
| based on ziya | 0.5759 | 0.6028 |
| based on chinese_alpaca-7b | 0.5771 | 0.6063 |
| average Acc | 0.5761 | 0.5994 |

Table 5-2 Responsibility Evaluation Results

By using the CValues evaluation method and setting different large models for a comprehensive assessment of the Responsibility Acc metric, it can be seen that further training alignment learning on the safety-aligned dataset we constructed has also led to an improvement in Responsibility Acc.

Finally, we assessed the model's general capabilities, and the benchmark evaluation results are shown in Table 5-3:

| Eval Set | Llama3.2-1B-Instruct | Llama3.2-1B-Instruct-ours | Gain |
|---|---|---|---|
| gsm8k | 44.4 | 43.90 | -0.50 |
| mmlu | 45.60 | 45.53 | -0.07 |
| cmmlu | 37.48 | 37.49 | +0.01 |
| ceval | 37.96 | 40.49 | +2.53 |
| HumanEval | 38.41 | 38.41 | +0.00 |

Table 5-3 Benchmark Evaluation Results

Through the evaluation and analysis of the model's general capability benchmark indicators, it can be observed that after further training alignment learning on the safety-aligned dataset we constructed, there is no significant decline in the benchmark indicators, and there is a significant improvement in ceval.

## 6 Conclusion

In this paper, we propose a multi-dimensional attack-defense alignment data construction method, which is an effective method for constructing safety instruction alignment data. By using the instruction alignment data constructed with our method, we conducted comparative experiments based on open-source large models and designed a more comprehensive safety assessment benchmark and evaluation method. The test results show an approximate increase of 4 percentage points in SafetyBench and an average increase of over 30% in various security scenarios in the more comprehensive evaluation benchmark tests. These test results demonstrate that the safety alignment learning based on the method proposed in this paper can significantly enhance the security defense effect of large models, while also maintaining and improving the model's general capabilities. Finally, although this paper only verifies the improvement of Chinese safety, the method is also universal and applicable to other languages.

## References

[1] Mireshghallah, Niloofar et al. "Can LLMs Keep a Secret? Testing Privacy Implications of Language Models via Contextual Integrity Theory." ArXiv abs/2310.17884 (2023): n. pag.
[2] Chen, Canyu and Kai Shu. "Can LLM-Generated Misinformation Be Detected?" ArXiv

abs/2309.13788 (2023): n. pag.

**[3]** Anwar, Usman et al. "Foundational Challenges in Assuring Alignment and Safety of Large Language Models." ArXiv abs/2404.09932 (2024): n. pag.

**[4]** Chung, Hyung Won et al. "Scaling Instruction-Finetuned Language Models." ArXiv abs/2210.11416 (2022): n. pag.

**[5]** Ouyang, Long et al. "Training language models to follow instructions with human feedback." ArXiv abs/2203.02155 (2022): n. pag.

**[6]** Bianchi, Federico et al. "Safety-Tuned LLaMAs: Lessons From Improving the Safety of Large Language Models that Follow Instructions." ArXiv abs/2309.07875 (2023): n. pag.

**[7]** Rafailov, Rafael, Archit Sharma, Eric Mitchell, Stefano Ermon, Christopher D. Manning and Chelsea Finn. "Direct Preference Optimization: Your Language Model is Secretly a Reward Model." ArXiv abs/2305.18290 (2023): n. pag.

**[8]** Ethayarajh, Kawin, Winnie Xu, Niklas Muennighoff, Dan Jurafsky and Douwe Kiela. "KTO: Model Alignment as Prospect Theoretic Optimization." ArXiv abs/2402.01306 (2024): n. pag.

**[9]** Ji, Jiaming, et al. "Beavertails: Towards improved safety alignment of llm via a human-preference dataset." Advances in Neural Information Processing Systems 36 (2024).

**[10]** Dai, Josef et al. "Safe RLHF: Safe Reinforcement Learning from Human Feedback." ArXiv abs/2310.12773 (2023): n. pag.

**[11]** Chen, Sizhe et al. "Aligning LLMs to Be Robust Against Prompt Injection." ArXiv abs/2410.05451 (2024): n. pag.

**[12]** Chuang, Yung-Sung et al. "DoLa: Decoding by Contrasting Layers Improves Factuality in Large Language Models." ArXiv abs/2309.03883 (2023): n. pag.

**[13]** Li, Jianwei and Jung-Eun Kim. "Superficial Safety Alignment Hypothesis." ArXiv abs/2410.10862 (2024): n. pag.

**[14]** Li, Shen et al. "Safety Layers in Aligned Large Language Models: The Key to LLM Security." (2024).

**[15]** Deng, Yue et al. "Multilingual Jailbreak Challenges in Large Language Models." ArXiv abs/2310.06474 (2023): n. pag.

**[16]** Zhang, Zhexin et al. "SafetyBench: Evaluating the Safety of Large Language Models with Multiple Choice Questions." *ArXiv* abs/2309.07045 (2023): n. pag.

**[17]** Xu, Guohai et al. "CValues: Measuring the Values of Chinese Large Language Models from Safety to Responsibility." *ArXiv* abs/2307.09705 (2023): n. pag.

**[18]** Yuan, Xiaohan et al. "S-Eval: Automatic and Adaptive Test Generation for Benchmarking Safety Evaluation of Large Language Models." *ArXiv* abs/2405.14191 (2024): n. pag.

**[19]** Zhai, Keke. "A Post-Training Enhanced Optimization Approach for Small Language Models." (2024)

**[20]** Sun, Hao, Zhexin Zhang, Jiawen Deng, Jiale Cheng and Minlie Huang. "Safety Assessment of Chinese Large Language Models." *ArXiv* abs/2304.10436 (2023): n. pag.

**[21]** https://www.modelscope.cn/datasets/iic/CValues-Comparison

**[22]** https://llmbench.ai/safety

**[23]** https://github.com/X-PLUG/CValues

**[24]** https://github.com/hiyouga/LLaMA-Factory

**[25]** https://github.com/QwenLM/Qwen/tree/main/eval

**[26]** https://github.com/openai/human-eval